\documentclass[aps,prc,twocolumn,superscriptaddress,showpacs,floatfix]{revtex4-1}
\usepackage{graphicx,amssymb}
\usepackage[fleqn]{amsmath}
\usepackage{amsfonts}
\usepackage{dcolumn}
\usepackage{bm}
\usepackage{braket}
\usepackage{multirow} 
\usepackage{MnSymbol}

\usepackage{CJKutf8}

\graphicspath{{images/}}

\newcommand{\hatvec}[1]
{\hat{\vec{#1}}}

\renewcommand{\vec}[1]{\mbox{\boldmath $#1$}}

\begin{document}
\begin{CJK*}{UTF8}{gbsn}

\date{\today}

\title{Single-particle and collective motion in unbound deformed ${ {}^{39}\text{Mg} }$}

\author{K. Fossez}
\affiliation{NSCL/FRIB Laboratory,
Michigan State University, East Lansing, Michigan 48824, USA}

\author{J. Rotureau}
\affiliation{NSCL/FRIB Laboratory,
Michigan State University, East Lansing, Michigan 48824, USA}
\affiliation{JINPA, Oak Ridge National Laboratory, Oak Ridge, TN 37831, USA}

\author{N. Michel}
\affiliation{NSCL/FRIB Laboratory,
Michigan State University, East Lansing, Michigan 48824, USA}

\author{Quan Liu (刘泉)}
\affiliation{NSCL/FRIB Laboratory,
Michigan State University, East Lansing, Michigan 48824, USA}
\affiliation{School of Physics and Material Science, Anhui University, Hefei 230601, People’s Republic of China}

\author{W. Nazarewicz}
\affiliation{Department of Physics and Astronomy and NSCL/FRIB Laboratory,
Michigan State University, East Lansing, Michigan 48824, USA}
\affiliation{Institute of Theoretical Physics, Faculty of Physics, 
University of Warsaw, Warsaw, Poland}

\begin{abstract}
	\begin{description}
		\item[Background]
			Deformed neutron-rich magnesium isotopes constitute a fascinating territory where the interplay between collective rotation and single-particle motion is strongly affected by the neutron continuum. The unbound ${ fp }$-shell nucleus ${ {}^{39}\text{Mg} }$ is an ideal candidate to study this interplay.
		\item[Purpose]
			In this work, we predict the properties of low-lying resonant states of ${ {}^{39}\text{Mg} }$, using a suite of realistic theoretical approaches rooted in the open quantum system framework.
		\item[Method]
			To describe the spectrum and decay modes of ${ {}^{39}\text{Mg} }$ we use the conventional Shell Model, Gamow Shell Model, Resonating Group Method, Density Matrix Renormalization Group method, and the non-adiabatic Particle-Plus-Rotor model formulated in the Berggren basis.
		\item[Results]
			The unbound ground state of ${ {}^{39}\text{Mg} }$ is predicted to be either a ${ {J}^{ \pi } = {7/2}^{-} }$ state or a ${ {3/2}^{-} }$ state. A narrow ${ {J}^{ \pi } = {7/2}^{-} }$ ground-state candidate exhibits a resonant structure reminiscent of that of its one-neutron halo neighbor ${ {}^{37}\text{Mg} }$, which is dominated by the ${ {f}_{7/2} }$ partial wave at short distances and a ${ {p}_{3/2} }$ component at large distances. A ${ {J}^{ \pi } = {3/2}^{-} }$ ground-state candidate is favored by the large deformation of the system. It can be associated with the ${ {1/2}^{-} [321] }$ Nilsson orbital dominated by the $\ell=1$ wave; hence its predicted width is large. 
			The excited ${J}^{ \pi } = {1/2}^{-}$ and $5/2^-$ states are expected to be broad resonances, 
			while the ${ {J}^{ \pi } = {9/2}^{-} }$ and ${ {11/2}^{-} }$ members of the ground-state rotational band are predicted to have very small neutron decay widths.
		\item[Conclusion]
			We demonstrate that the subtle interplay between deformation, shell structure, and continuum coupling can result in a variety of excitations in an unbound nucleus just outside the neutron drip line.
	\end{description}
\end{abstract}


\maketitle
\end{CJK*}

\section{Introduction}

Weakly bound and unbound nuclei in the vicinity of the neutron drip line offer a unique window into the interplay between single particle (s.p.) degrees of freedom, collective motion, and a variety of couplings to the continuum of scattering and decay channels. 
While many phenomena found in systems close to the neutron or two-neutron drip lines, such as the appearance of halo structures \cite{jensen04_233,tanihata13_549}, clustering effects \cite{oertzen06_1017,freer07_1018,okolowicz13_241,okolowicz12_998}, many-neutron correlations \cite{kisamori16_1463,pieper03_1461}, and di-neutron radioactivity \cite{kohley15_1213,kondo16_1439,hagino14_1160}, are general to all open quantum systems, every new case sheds light on intricate threshold couplings governing the existence of nuclei on the verge of particle stability.

The region of ``inversion'' in the neutron-rich territory between the conventional magic gaps ${ N = 20 }$ and ${ N = 28 }$ is characterized by large quadrupole deformations rich in phenomena associated with shape coexistence \cite{heyde11_1483,gade16_1606}.
The case of unbound ${ {}^{39}\text{Mg} }$ \cite{baumann07_1454,watanabe14_1449} is of particular interest in this context, because of the unique position of this nucleus in the nuclear landscape. Indeed, ${ {}^{39}\text{Mg} }$ lies at the neutron-rich side of the Mg chain, just one neutron away from the deformed ${ N = 28 }$ isotope
${ {}^{40}\text{Mg} }$ \cite{crawford14_1607}.
It is one of the heaviest neutron-rich unbound systems experimentally available before the next generation of rare isotope beam facilities comes online \cite{thoennessen04_1165}; hence, it represents a bridge between the well known halo structures in the light ${ psd }$ nuclei and the suspected halos in medium-mass nuclei \cite{rotival09_1587,duguet16_1588}.

From a theoretical point of view, the description of ${ {}^{39}\text{Mg} }$ is challenging because of two interrelated aspects of this open quantum system. First, the even-even magnesium isotopes 
${ {}^{36,38,40}\text{Mg} }$ are relatively weakly bound \cite{sakurai97_1445,crawford14_1607} and are predicted to have a similar quadrupole deformation of about ${ { \beta }_{2} = 0.3 }$ \cite{nowacki09_1450,caurier14_1509,erler12_1297,massexplorer}. The addition of a neutron to 
${ {}^{36}\text{Mg} }$ leads to the one-neutron halo nucleus ${ {}^{37}\text{Mg} }$ \cite{kobayashi14_1446,takechi14_1448}; hence, a ${ p }$-wave dominated structure is also expected in ${ {}^{39}\text{Mg} }$, but just above the one-neutron threshold. This particular situation is of interest for the studies of ``deformed'' halo systems, such as ${ {}^{17}\text{C} }$, ${ {}^{31}\text{Ne} }$, and ${ {}^{37}\text{Mg} }$ \cite{misu97_1181,zhou10_1572}, since ${ {}^{39}\text{Mg} }$ can provide an insight on how deformed halo structures evolve as the neutron chemical potential dives into the particle continuum.

Second, the competition between the ${ s }$ and ${ d }$ partial waves known from the one-neutron halo system ${ {}^{11}\text{Be} }$ translates into a competition between the ${ f }$ and ${ p }$ shells, which results in
the unbound ground state (g.s.) of ${ {}^{39}\text{Mg}}$. In both cases, an odd-${ A }$ nucleus can be viewed in terms of a neutron coupled to a deformed core. Recently,  within a model that can account for the deformed character of the core and the 
valence-particle continuum,  we demonstrated that long-lived unbound collective states can in fact exist in ${ {}^{11}\text{Be} }$ at high excitation energy \cite{fossez16_1335}. Whether a similar situation can occur in a heavier ${ {}^{39}\text{Mg} }$, where the notion of collective motion is better founded, is an interesting question.

This article is organized as follows: Section \ref{sec_models_and_parameters} contains the description of the models and methods used. In particular, it discusses the advantages and disadvantages of each method for the description of ${ {}^{39}\text{Mg} }$ and describes the effective interactions used as well as the related optimization procedure. Our structural predictions are contained in Sec.~\ref{sec_results}. Finally, the summary and outlook are given in Sec.~\ref{sec_conclusion}.

\section{Models and parameters}\label{sec_models_and_parameters}

The theoretical description of ${ {}^{39}\text{Mg} }$ is demanding. The presence of large quadrupole deformations in the neutron-rich magnesium isotopes, complex many-body dynamics and resulting shell evolution, and a pronounced coupling to the neutron continuum, all require the use of diverse approaches. In this study, we use the standard nuclear shell model (SM); the Gamow shell model (GSM) \cite{michel09_2}; the open-system extension of the density matrix renormalization group method (DMRG) \cite{rotureau06_15,rotureau09_140}; the GSM in the resonating group method formalism (RGM) \cite{jaganathen14_988,fossez15_1119}; and the non-adiabatic particle-plus-rotor model (PRM) solved by using the Berggren
expansion method \cite{fossez13_552,fossez15_1028,fossez16_1335}.

The deformation of the core is explicitly included only in the PRM through the description of ${ {}^{38}\text{Mg} }$ by a quadrupole-deformed Woods-Saxon (WS) potential. In other approaches, the deformation enters indirectly through the configuration mixing.
Except for the SM, the continuum is explicitly included using the Berggren expansion method, but at different levels: while the PRM considers only one neutron in the positive-energy continuum, GSM-based approaches allow several particles to scatter into unbound levels.

\subsection{Berggren basis}\label{sec_Berggren_basis}

The s.p. Berggren ensemble is defined through the completeness relation including explicitly the Gamow (resonant) states and the non-resonant scattering continuum \cite{berggren68_32,berggren82_28}. For each considered partial wave ${ c = ( \ell , j ) }$
the completeness relation takes the form 
	\begin{equation}
		\sum_{i} \ket{ {u}_{c} ( {k}_{i} ) } \bra{ \tilde{u}_{c} ( {k}_{i} ) } + \int_{ \mathcal{L}_{c}^{+} } dk \, \ket{ {u}_{c} (k) } \bra{ \tilde{u}_{c} (k) } = \hat{1}_{c},
		\label{eq_Berggren_basis}
	\end{equation}
	where ${ \ket{ {u}_c ( {k}_{i} ) } }$ are radial wave functions of the resonant states and ${ \ket{ {u}_{c} (k) } }$ are complex-energy scattering states along a contour ${ \mathcal{L}_{c}^{+} }$ in the complex-momentum plane. In our applications, ${ \mathcal{L}_{c}^{+} }$ starts at zero, surrounds the selected resonant states in the fourth quadrant of the complex-momentum plane, and extends to ${ k \to +\infty }$ to ensure  completeness. The tilde in Eq.~\eqref{eq_Berggren_basis} means \textit{time-reversed}, which translates into the complex conjugation for unbound states. Thanks to Cauchy's integral theorem, the precise form of the contour ${ \mathcal{L}_{c}^{+} }$ is unimportant, provided that all the selected discrete states lie between the contour and the real-${ k }$ axis. The selection of the discrete states entering in Eq.~\eqref{eq_Berggren_basis} depends on the problem in hand.

The s.p. bound states entering Eq.~\eqref{eq_Berggren_basis} are normalized in the usual way, while the decaying resonant states require the use of the exterior complex scaling method \cite{dykhne61_1041,gyarmati71_38}. The non-resonant states are normalized to the Dirac delta distribution. In practice, the integral along the contour ${ \mathcal{L}_{c}^{+} }$ in Eq.~\eqref{eq_Berggren_basis} is discretized using Gauss-Legendre quadrature, and the selected scattering states are renormalized by the quadrature weights, which results in the Kronecker delta normalization.

In the following, the contour is always defined by three segments, with the first one starting at the origin and ending at ${ {k}_{ \text{peak} } = \alpha - i \beta }$ with ${ \alpha, \beta > 0 }$; the second one going from ${ {k}_{ \text{peak} } }$ to ${ {k}_{ \text{middle} } }$ (real); and the last one starting at ${ {k}_{ \text{middle} } }$ and ending at ${ {k}_{ \text{max} } }$ (real). The momentum cutoff ${ {k}_{ \text{max} } }$ has to be sufficiently large to ensure completeness.

\subsection{Models}\label{sec_models}

The GSM is the SM formulated in the complex-energy plane \cite{michel09_2} via the replacement of the usual harmonic oscillator (HO) s.p. basis by the Berggren basis. While the GSM Hamiltonian is Hermitian, the Hamiltonian matrix is complex-symmetric, leading to complex eigenvalues. The complex-energy approach enables a general treatment of couplings between bound and unbound states of the many-body system rooted in a quasi-stationary picture.

In the SM, spurious center-of-mass excitations are removed using the Lawson method \cite{lawson80_b122,whitehead77_867}. In the GSM this method does not apply, because Berggren wave functions are not eigenstates of the HO potential. To circumvent the center of mass effect, the GSM Hamiltonian is expressed in the intrinsic nucleon-core coordinate of the cluster-orbital SM \cite{suzuki88_595}:
	\begin{equation}
		\hat{H} = \sum_{i = 1}^{ {N}_{ \text{val} } } \left( \frac{ \hatvec{p}_{i}^{2} }{ 2 { \mu }_{i} } + {U}_{c} ( \hat{r}_{i} ) \right) + \sum_{i < j}^{ {N}_{ \text{val} } } \left( V ( \hatvec{r}_{i} - \hatvec{r}_{j} ) + \frac{ {\hatvec{p}_{i}}{\cdot} {\hatvec{p}_{j} }}{ {M}_{c} } \right),
		\label{eq_GSM_Hamiltonian_COSM}
	\end{equation}
where ${ {N}_{ \text{val} } }$ is the number of valence nucleons, ${ {M}_{c} }$ is the mass of the core nucleus, and ${ 1/{ { \mu }_{i} } = 1/{ {M}_{c} } + 1/{ {m}_{i} } }$ is the reduced mass of a valence nucleon. The s.p. core potential acting on valence nucleons, ${ {U}_{c} ( \hat{r} ) }$, is a sum of the nuclear and Coulomb terms. The nuclear term is given by a WS potential with a spin-orbit term \cite{michel03_10}, while the Coulomb potential is generated by the Gaussian-shaped density of the ${ {Z}_{c} }$ protons of the core \cite{michel03_10}. In the present work, the two-body interaction ${ V ( \hatvec{r}_{i} - \hatvec{r}_{j} ) }$ is taken as the Furutani-Horiuchi-Tamagaki (FHT) finite-range two-body interaction \cite{furutani78_1012,furutani79_1013}, which contains nuclear and Coulomb terms (see Ref.~\cite{fossez15_1119} for a recent application).

In the GSM, the inherent configuration mixing makes the identification of physical decay channels difficult. To overcome this difficulty, the GSM can be formulated in the RGM representation~\cite{fossez15_1119}, where incoming and outgoing decay channels are explicit components of wave functions. In the RGM picture, the ``target'' as well as the ``projectile'' are described in the GSM, and the structure of the total system is obtained by solving the coupled-channel equations.
	
The RGM variant of the GSM can thus be viewed as an alternative to the continuum shell model and the shell model embedded in the continuum \cite{barz71_669,okolowicz03_21,volya06_94}, and can be used to describe both the nuclear structure and reactions in a unified framework. Within this approach the many-body correlations in the target are all included, at least up to the model space considered, and the Schr\"odinger equation formulated in the coupled-channel formalism is solved via direct integration, which means that the target-projectile continuum is treated exactly.

Another way to solve the nuclear many-body problem in the continuum is the DMRG approach \cite{white92_488,rotureau06_15,rotureau09_140}, which relies on the fact that in most cases the non-resonant continuum plays a perturbative role in shaping the many-body eigenstates of the Hamiltonian. This method constitutes a powerful truncation scheme to tame the dramatic increase of the Hilbert space in both the GSM and the no-core GSM \cite{papadimitriou13_441}. In DMRG, instead of constructing the full Hamiltonian matrix ${ H }$ as in the GSM, one considers a zero order approximation ${ {H}_{0} }$ of the full problem, obtained by considering the truncated Berggren basis that only contains discrete resonant states (poles). The eigenstates of ${ {H}_{0} }$ form the reference space and the remaining many-body states containing contributions from non-resonant s.p. states form the complement space. The eigenstates of ${ H }$ are obtained by adding the continuum degrees of freedom gradually, as dictated by the density matrix in the complement space.

Recently, the DMRG approach has been augmented by the use of natural orbitals for the \textit{ab initio} description of ${ {}^{6}\text{Li} }$ \cite{arxivShin16}. This technique significantly improves convergence with the number of included shells and is thus suitable when a large model space has to be considered. The DMRG is thus the tool of choice to extend GSM calculations into a region where a direct diagonalization of the Hamiltonian matrix is not feasible.

Another complementary approach used in this work is the non-adiabatic PRM solved in the Berggren basis \cite{fossez16_1335}. In the PRM picture, the core is described by a deformed WS potential with a spin-orbit term, and the continuum of the valence particle is fully taken into account by expanding the channel wave functions in the Berggren basis or by directly integrating the coupled channel equations. This model is a simplification of RGM, with the collectivity explicitly included. However, since a rigid core is assumed, the many-body dynamics related to the core is neglected in the PRM, and the core-particle antisymmetrization is treated approximately.

The two classes of methods, i.e., structure-oriented (SM, GSM, DMRG) and decay/reaction-oriented (RGM, PRM), form the backbone of the present theoretical description of ${ {}^{39}\text{Mg} }$. They allow for a comprehensive study of spectroscopy and decay channels of ${ {}^{39}\text{Mg} }$ that includes s.p. and collective effects in a consistent way. The feasibility of this approach is based on the assumptions that the deformed magnesium isotopes ${ {}^{34-40}\text{Mg} }$ are bound, and that in the cases of ${ {}^{37,39}\text{Mg} }$ there is an approximate separation of scale between the energy associated with the rotational motion of the bound core and the s.p. energy of the valence neutron. For that reason, the core potential and the 2-body interaction used are taken from a SM-based optimization for all the magnesium isotopes considered. The continuum is taken into account in the GSM, RGM, and  DMRG calculations for ${ {}^{37,39}\text{Mg} }$. An agreement between our configuration interaction (CI) models and the collective PRM, whose Hamiltonian is fitted independently, is a necessary condition for validating our assumptions.

\subsection{Optimized interactions}\label{sec_fit_FHT}

To make reliable predictions for neutron resonances, reliable input is needed as resonance widths are sensitive to the threshold energy. The optimization of the GSM interaction has been performed by considering the doubly-magic ${ {}^{28}\text{O} }$ core, with 4 valence protons and from 2 to 8 valence neutrons (${ {}^{34-40}\text{Mg} }$). While a ${ {}^{28}\text{O} }$ core may not be suitable when considering a small number of valence nucleons, one may expect this choice to be justified for larger numbers of valence particles. The model space is defined by the ${ 0{d}_{5/2} }$, ${ 1{s}_{1/2} }$, and ${ 0{d}_{3/2} }$ HO shells (poles) for the protons, and the ${ 0{f}_{7/2} }$, ${ 1{p}_{3/2} }$, ${ 0{f}_{5/2} }$, and ${ 1{p}_{1/2} }$ HO shells (poles) for the neutrons, with 5 additional HO neutron shells above ${ 0{f}_{7/2} }$, and 4 additional HO neutron shells above ${ 1{p}_{3/2} }$, ${ 0{f}_{5/2} }$, and ${ 1{p}_{1/2} }$. In order to reduce the size of the model space, truncations on the number of particle-hole excitations were imposed. Up to four particle-hole excitations were allowed within the pole shells, and only one neutron at a time was allowed to access the higher HO shells. The ${ {}^{28}\text{O} }$ core is described as a WS potential defined separately for protons and neutrons, which was optimized together with the two-body interaction. For feasibility reasons, the optimization of the GSM interaction was carried out at the SM level, i.e., the continuum coupling was neglected during the fitting process. This assumption is expected to primarily impact low-${ J }$ states, which contain large contributions from low-${ \ell }$ partial waves.

Among the parameters involved in the interaction fit 
are the depth ${ {V}_{0} }$ and the spin-orbit strength ${ {V}_{ \text{so} } }$ of the one-body proton and neutron WS potentials. The diffuseness and radius of the WS potentials have been fixed at ${ a = 0.65 \, \text{fm} }$ and ${ {R}_{0} = 3.85 \, \text{fm} }$, respectively, 
and the charge radius has been set to ${ {R}_{ \text{c} } = 3.85 \, \text{fm} }$.
The remaining parameters involved in the fit are the eight parameters of the FHT interaction, 
including all the spin and isospin components of the central, spin-orbit, and tensor parts of the interaction.
This leads to a total of 12 parameters.

The optimization was carried out for two cases. In the first variant (V1), we assumed the g.s. of ${ {}^{37}\text{Mg} }$ and ${ {}^{39}\text{Mg} }$ to be a  ${ {J}^{ \pi } = {3/2}^{-} }$ state and  made no assumption about the g.s. of ${ {}^{35}\text{Mg} }$. This choice corresponds to a situation where the g.s. of ${ {}^{37}\text{Mg} }$ and ${ {}^{39}\text{Mg} }$ are similar and dominated by ${ p }$-waves, as suggested in Refs.~\cite{kobayashi14_1446,takechi14_1448}. In order to obtain a reasonable quadrupole coupling, the energies of the ${ {J}^{ \pi } = {0}^{+}_{1} }$ and ${ {2}^{+}_{1} }$ states in ${ {}^{34,36,38}\text{Mg} }$ have been included in the fit, as well as the g.s. of ${ {}^{40}\text{Mg} }$. Since the masses of ${ {}^{38,39,40}\text{Mg} }$ are not known experimentally, for these nuclei we used the ${ {S}_{n} }$ values obtained from systematic trends as quoted in Ref.~\cite{ensdf}.
In the second variant (V2), we assumed the g.s. of ${ {}^{35,37,39}\text{Mg} }$ to have ${ {J}^{ \pi } = {7/2}^{-} }$.

\begin{figure}[htb]
		\includegraphics[width=1.0\linewidth]{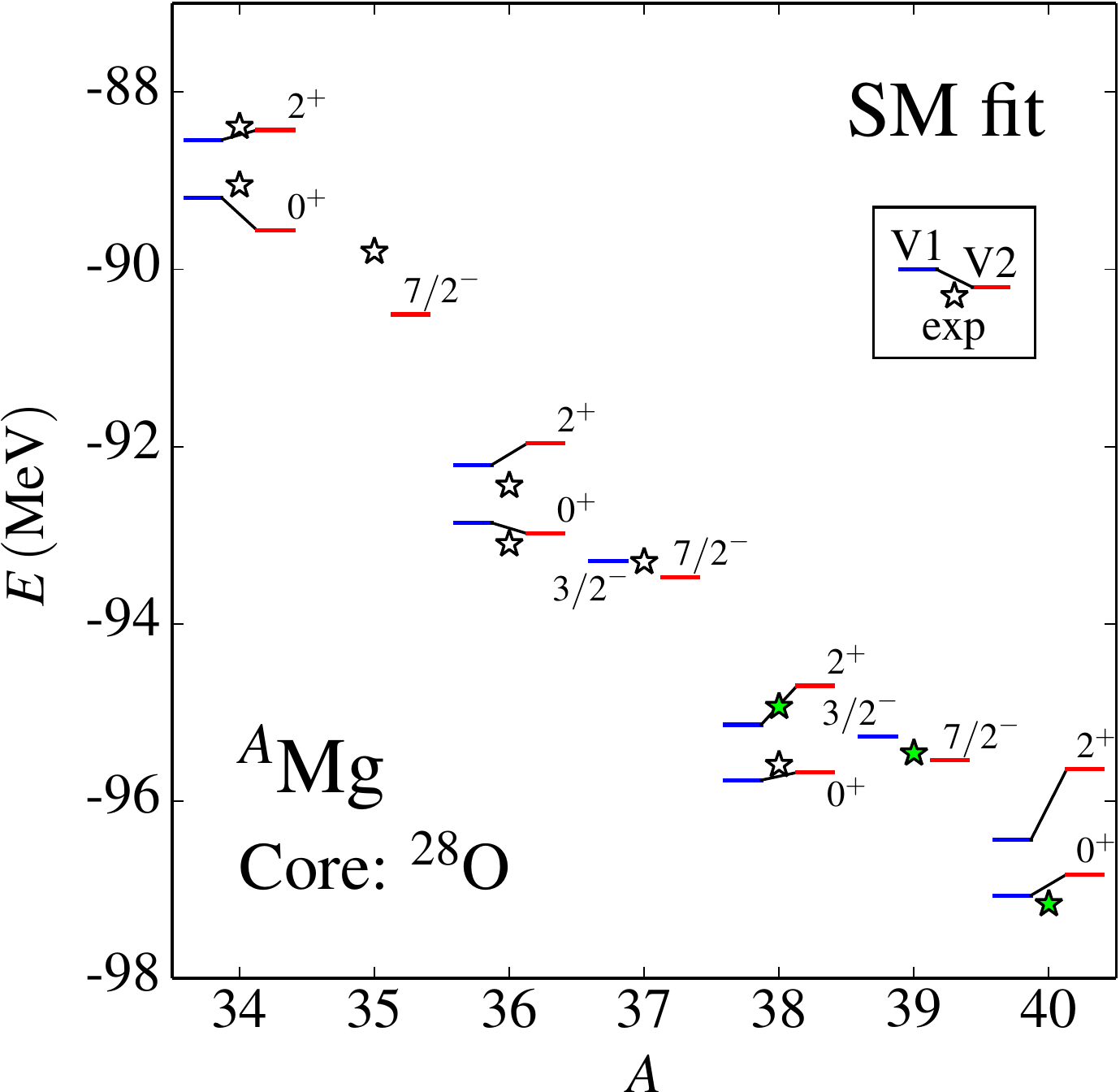}
		\caption{Fit of the GSM effective interaction for ${ {}^{34-40}\text{Mg} }$ carried out at the SM level. 
		Experimental data \cite{ensdf} are denoted by stars. Since the masses of ${ {}^{38,39,40}\text{Mg} }$ have not yet been measured, for these nuclei we used the ${ {S}_{n} }$ values (marked by filled stars) obtained from systematic trends, as quoted in Ref.~\cite{ensdf}.
			The results of the optimization variants V1 and V2 are marked by blue and red lines, respectively.}
		\label{fig_1}
	\end{figure}

The first variant V1 of the fit gives the depths and spin-orbit terms of the proton and neutron core potentials 
${ {V}_{0}^{ \text{p} } = -67.5 \, \text{MeV} }$, ${ {V}_{ \text{so} }^{ \text{p} } = 10.1 \, \text{MeV} }$, 
${ {V}_{0}^{ \text{n} } = -42.9 \, \text{MeV} }$ and ${ {V}_{ \text{so} }^{ \text{n} } = 5.96 \, \text{MeV} }$.
These parameters are similar to the second variant V2 of the fit: 
${ {V}_{0}^{ \text{p} } = -67.3 \, \text{MeV} }$, ${ {V}_{ \text{so} }^{ \text{p} } = 10.0 \, \text{MeV} }$, 
${ {V}_{0}^{ \text{n} } = -38.1 \, \text{MeV} }$ and ${ {V}_{ \text{so} }^{ \text{n} } = 10.0 \, \text{MeV} }$, 
with a slightly lower depth in the WS potential for neutrons.
The values of the parameters of the FHT interaction are given in Table~\ref{tab_FHT_params} for both fits, 
and the optimization outcome is shown in Fig.~\ref{fig_1}.

	\begin{table}[htb]
		\caption{Strengths of the central (c), spin-orbit (so), and tensor (t) terms of the FHT interaction in different spin-isospin channels obtained in the V1 and V2 variants. The values of
			${ {V}_{ \text{c} }^{S,T} }$ and ${ {V}_{ \text{so} }^{S,T} }$ are in MeV and ${ {V}_{ \text{t} }^{S,T} }$ is in ${ \text{MeV} \, \text{fm}^{-2} }$.
		The central term ${ {V}_{\text{c}}^{0,1} = -0.08 \, \text{MeV} }$ and spin-orbit term ${ {V}_{\text{so}}^{1,0} = 0.0 \, \text{MeV} }$ are identical in both fits.}
		\begin{ruledtabular}
			\begin{tabular}{lcccccc}
				Fit & ${ {V}_{\text{c}}^{1,1} }$ & ${ {V}_{\text{c}}^{1,0} }$ & ${ {V}_{\text{c}}^{0,0} }$ & ${ {V}_{\text{so}}^{1,1} }$ & ${ {V}_{\text{t}}^{1,1} }$ & ${ {V}_{\text{t}}^{1,0} }$ \\[2pt]
				\hline\\[-7pt]
				V1 & -3.38 & -2.74 & 3.48 & -78.9 & 21.4 & 31.0 \\
				V2 & -3.93 & -3.65 & 2.27 & -76.2 & 37.6 & 15.7 \\
			\end{tabular}
		\end{ruledtabular}
		\label{tab_FHT_params}
	\end{table}

	In both cases a reasonable agreement with experimental data is obtained. The fit V1 yields a slightly better description of the even-even isotopes than V2 but still predicts the g.s. of ${ {}^{37,39}\text{Mg} }$ to be ${ {J}^{ \pi } = {7/2}^{-} }$ states. 
	This is because the ${ 1{p}_{3/2} }$ neutron shell lies above the ${ 0{f}_{7/2} }$ shell and the ${ {p}_{3/2} }$ continuum is neglected in the SM description. 
	Since V2 provides a better description of ${ {S}_{n} }$ in ${ {}^{37,39}\text{Mg} }$, and -- as discussed below -- the continuum coupling plays a major role in description of the ${ {3/2}^{-} }$ state, in the following we choose to use this interaction to study the spectrum of ${ {}^{39}\text{Mg} }$.

\subsection{Optimization of the deformed PRM potential}\label{sec_fit_WS}

In the PRM, the interaction between the valence neutron and the effective core is represented by a deformed WS potential. The range of the WS parameters, i.e., the diffuseness ${ a }$, radius ${ {R}_{0} }$, strength ${ {V}_{0} }$, and the spin-orbit strength ${ {V}_{ \text{so} } }$, spans a range of ${ a = 0.67 \pm 6\% \, \text{fm} }$, ${ {R}_{0} = 4.35 \pm 11.5\% \, \text{fm} }$, ${ {V}_{0} = 37.0 \pm 5.5\% \, \text{MeV} }$, and ${ {V}_\text{so} = 8.1 \pm 5\% \, \text{MeV} }$. The quadrupole deformation of the WS potential is in the range ${ { \beta }_{2} = 0.3 \pm 0.02 (6.67\%) }$, which is consistent with the Hartree-Fock-Bogoliubov predictions of Refs.~\cite{nazarewicz01_1180,erler12_1297,massexplorer}. The g.s. of ${ {}^{37}\text{Mg} }$ and ${ {}^{39}\text{Mg} }$ are independently considered during the optimization process. In this way, the deformed WS potentials for both systems are consistent with each other and have reasonable quadrupole deformations.

The coupled-channel equations of the PRM \cite{fossez16_1335} were solved up to a maximal radius of ${ {R}_{ \text{max} } = 30 \, \text{fm} }$ and the rotation point for the exterior complex-scaling was fixed at ${ {R}_{ \text{rot} } = 20 \, \text{fm} }$. The valence space was assumed to consist of the ${ 0{f}_{7/2} }$, ${ 1{p}_{3/2} }$, ${ 0{f}_{5/2} }$, and ${ 1{p}_{1/2} }$ neutron resonant shells, which were augmented by the corresponding non-resonant partial waves. For each partial wave, the complex-energy scattering states entering the Berggren basis \eqref{eq_Berggren_basis} were selected along a contour ${ \mathcal{L}_{c}^{+} }$ defined in the complex-momentum plane by the points ${ (0,0) }$, ${ (0.2,-0.2) }$, ${ (0.5,0) }$ and ${ (3.5,0) }$ (all in ${ \text{fm}^{-1} }$). The contour was discretized with 70 points using a Gauss-Legendre quadrature. The rotational energies of the core nuclei were fixed at their experimental values \cite{ensdf}
of ${ E ({2}^{+}_{1}) = 0.660 \text{MeV} }$ and ${ E ({4}^{+}_{1}) = 2.016 \, \text{MeV} }$ for ${ {}^{36}\text{Mg} }$, and
${ E ({2}^{+}_{1}) = 0.656 \, \text{MeV} }$ and ${ E ({4}^{+}_{1}) = 2.016 \text{MeV} }$ for ${ {}^{38}\text{Mg} }$. For the energies of higher-lying g.s. band members, we assumed the moment of inertia corresponding to that of the ${ {4}^{+}_{1} }$ state. The resulting WS parameters are listed in Table~\ref{tab_defWS_params}.
	\begin{table}[htb]
		\caption{Parameters of the WS potentials for the ${ {}^{37}\text{Mg} }$ and ${ {}^{39}\text{Mg} }$ nuclei obtained in this work.}
		\begin{ruledtabular}
			\begin{tabular}{cccccc}
				Nucleus & ${ a }$ & ${ {R}_{0} }$ & ${ {V}_{0} }$ & ${ {V}_{ \text{so} } }$ & ${ { \beta }_{2} }$ \\
				& (fm) & (fm) & (MeV) & (MeV) & \\ 
				\hline \\[-6pt]
				$^{37}$Mg 	& 0.690 & 4.211 & -35.77 & 7.884 & 0.293 \\
				\\[-6pt]
				$^{39}$Mg 	& 0.728 & 4.206 & -35.00 & 8.000 & 0.300 
			\end{tabular}
		\end{ruledtabular}
		\label{tab_defWS_params}
	\end{table}
	For both nuclei, the predicted g.s. is ${ {J}^{ \pi } = {3/2}^{-} }$, with a ${ {J}^{ \pi } = {7/2}^{-} }$ level being very close in energy. As discussed below, 
	the absence of antisymmetry between the core and the valence neutron lowers the energy of the ${ {J}^{ \pi } = {3/2}^{-} }$ state in the PRM.

	\section{Low-energy structure of ${ {}^{39}\text{Mg} }$}\label{sec_results}
	
\subsection{Excitation spectrum}\label{sec_excited_states}

\begin{figure}[htb]
		\includegraphics[width=0.90\linewidth]{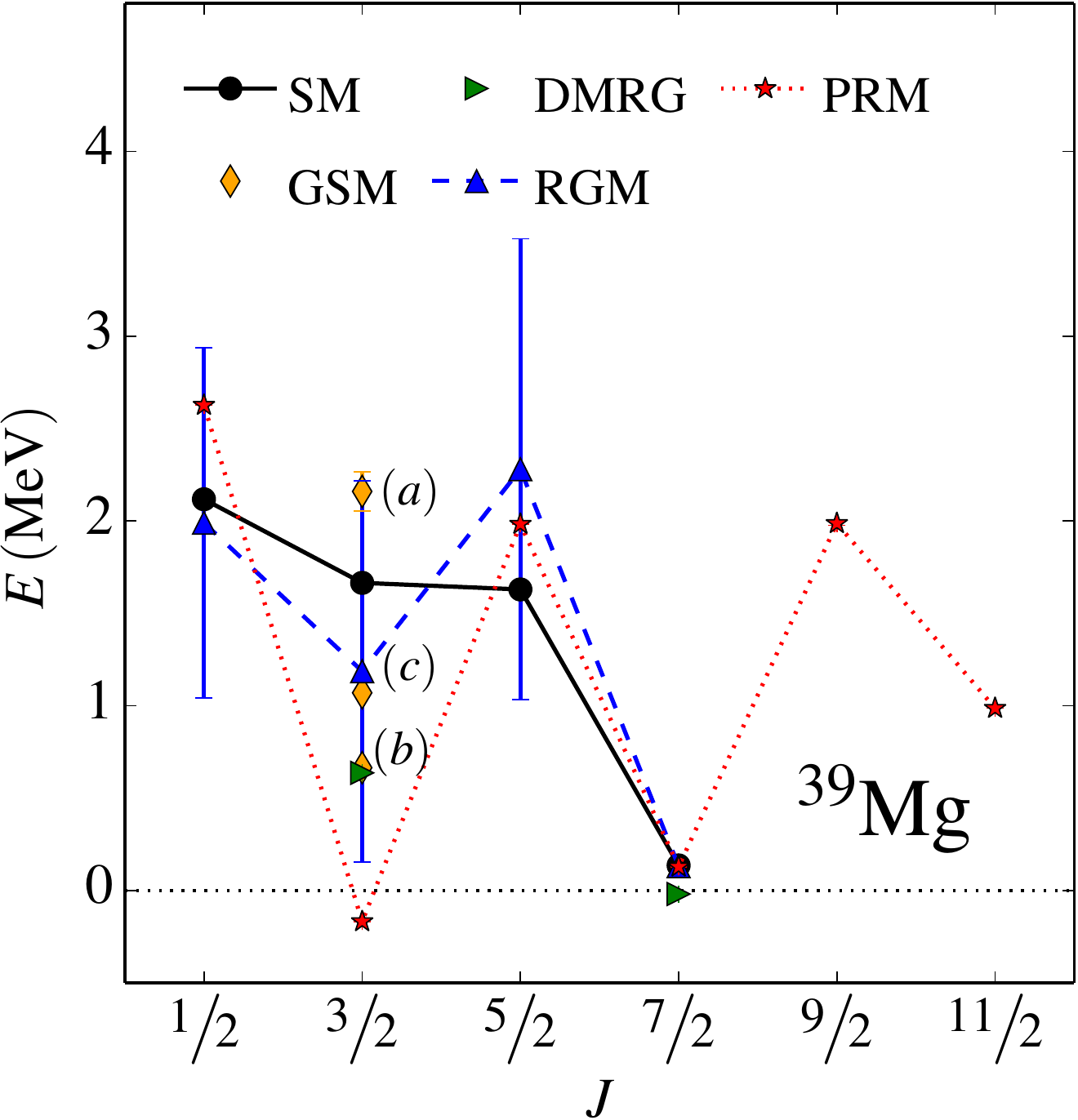}
		\caption{Excited states of ${ {}^{39}\text{Mg} }$ predicted in different models used in this work (see text for more details).}
		\label{fig_2}
	\end{figure}

The energies and widths of low-lying states of ${ {}^{39}\text{Mg} }$ predicted in different models are shown in Fig.~\ref{fig_2}. Only states with ${ J \leq 7/2 }$ are calculated using CI models (SM, GSM, RGM, DMRG) since the description of higher-${ J }$ states would require significantly larger model spaces.

\begin{figure}[htb]
		\includegraphics[width=0.90\linewidth]{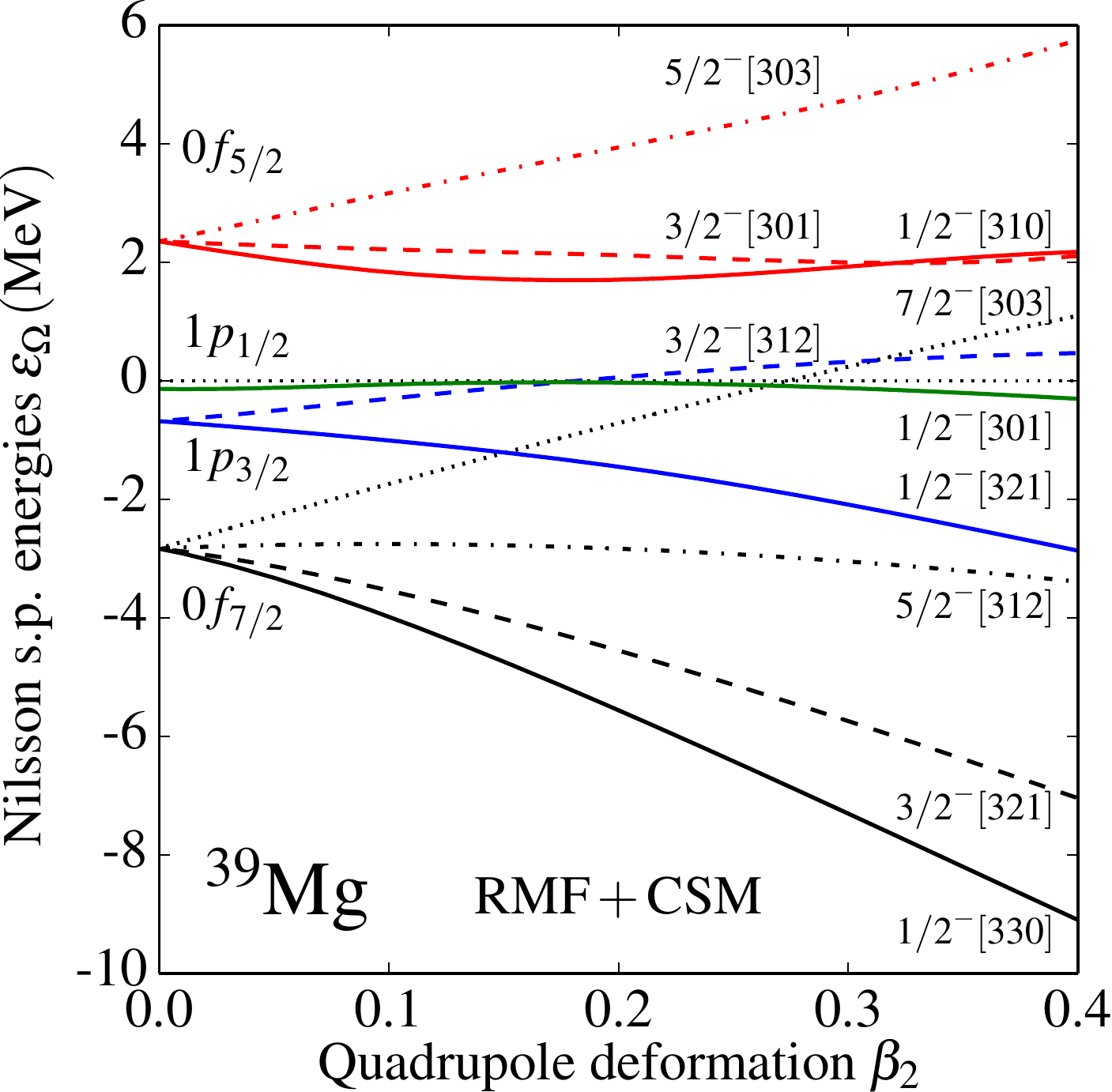}
		\caption{Single-particle neutron Nilsson diagram for ${ {}^{39}\text{Mg} }$ from the relativistic mean-field approach with the complex-scaling method. The crossing between the deformed levels
		${ {7/2}^{-} [303] }$ and ${ {1/2}^{-} [321] }$ originating from the ${ 0{f}_{7/2} }$ and ${ 1{p}_{3/2} }$ shells, respectively, results in a deformed subshell closure at ${ N = 28 }$ and ${ { \beta }_{2} \approx 0.3 }$.}
		\label{fig_3}
	\end{figure}

	The general properties of the level scheme are generally consistent across the models: the ${ {J}^{ \pi } = {1/2}^{-} }$ and ${ {5/2}^{-} }$ states lie higher in energy as compared to the ${ {3/2}^{-} }$ and ${ {7/2}^{-} }$ levels. To understand this pattern let us first discuss the PRM results, which have a simple interpretation in terms of the large deformation of the core and the underlying spectrum of deformed s.p. levels. Figure~\ref{fig_3} shows the Nilsson diagram obtained using the relativistic mean-field approach with the complex-scaling method \cite{guo10_1577,liu12_1576,liu13_1575}. (For other deformed s.p. level diagrams in this region, see Refs.~\cite{nazarewicz01_1180,hamamoto07_1511,hamamoto09_1452,hamamoto16_1609,watanabe14_1449,takechi14_1448,xu15_1586}.)
There are two features of the s.p. neutron spectrum that are very relevant to
properties of ${ {}^{39}\text{Mg} }$. First, the spherical ${ 0{f}_{7/2} }$, ${ 1{p}_{3/2} }$, and ${ 1{p}_{1/2} }$ shells appear close in energy due to the concentration 
of the low-${ \ell }$ s.p. strength around the zero-energy threshold~\cite{nazarewicz01_1180,hamamoto07_1511,hamamoto09_1452,hamamoto16_1609,xu15_1586}. This results in a dramatic quenching of the spherical ${ N = 28 }$ neutron gap. Second, as a result of the crossing between the Nilsson levels ${ {7/2}^{-} [303] }$ and ${ {1/2}^{-} [321] }$ a large deformed ${ N = 28 }$ subshell closure appears around ${ { \beta }_{2} = 0.3 }$. This gap supports the strongly deformed shape of ${ {}^{40}\text{Mg} }$ predicted by theory \cite{nazarewicz01_1180} and suggest the Nilsson model assignment ${ {1/2}^{-} [321] }$ for the g.s. of ${ {}^{39}\text{Mg} }$. Since the rotational decoupling parameter associated with the ${ {1/2}^{-} [321] }$ orbit is slightly below ${ a = -1 }$, one would expect a ${ {J}^{ \pi } = {3/2}^{-} }$ g.s. assignment according to the strong-coupling limit of the PRM~\cite{bohr98_b27}. This result is nicely consistent with the non-adiabatic PRM result in Fig.~\ref{fig_3}, which predicts the large signature splitting between the favored (${ r = \exp ( -i \pi J ) = i }$) and unfavored (${ r = -i }$) members of the g.s. band of of ${ {}^{39}\text{Mg} }$ as well as the presence of close-lying ${ {J}^{ \pi } = {3/2}^{-} }$ and ${ {7/2}^{-} }$ levels. 
\begin{table}[htb]
	\caption{Contributions to the norm of ${ {J}^{ \pi } }$ states (in percent) of ${ {}^{39}\text{Mg} }$ from the considered partial waves ${ ( \ell , j ) }$ in the PRM.}
		\begin{ruledtabular}
			\begin{tabular}{ccccc}
				${ {J}^{ \pi } }$ & ${ {n}_{ {f}_{7/2} } }$ & ${ {n}_{ {p}_{3/2} } }$ & ${ {n}_{ {f}_{5/2} } }$ & ${ {n}_{ {p}_{1/2} } }$ \\
				\hline \\[-6pt]
				${ {1/2}^{-} }$ & 13 & 85 & 0 & 2 \\
				${ {3/2}^{-} }$ & 32 & 62 & 1 & 5 \\
				${ {5/2}^{-} }$ & 12 & 12 & 1 & 75 \\
				${ {7/2}^{-} }$ & 62 & 36 & 0 & 2 \\
				${ {9/2}^{-} }$ & 62 & 30 & 3 & 4 \\
				${ {11/2}^{-} }$ & 80 & 19 & 0 & 1 
			\end{tabular}
		\end{ruledtabular}
		\label{tab_defWS_channel_norms}
	\end{table}

The relative importance of various partial waves in a given PRM state can be estimated from the norm of each channel wave function by summing over the core angular momentum. The resulting partial wave contributions to the norm are shown in Table~\ref{tab_defWS_channel_norms}
and reveal that the ${ {J}^{ \pi } = {1/2}^{-} }$ and ${ {3/2}^{-} }$ states have a large ${ {p}_{3/2} }$ component, the ${ {J}^{ \pi } = {5/2}^{-} }$ state has a large ${ {p}_{1/2} }$ component, and other states are mostly dominated by the ${ {f}_{7/2} }$ partial wave.

According to the PRM, the main components of the ${ {J}^{ \pi } = {7/2}^{-} }$ state are ${ \ket{ {0}^{+} } \otimes \ket{ {f}_{7/2} } }$ and 
${ \ket{ {2}^{+} } \otimes \ket{ {p}_{3/2} } }$ and those for ${ {J}^{ \pi } = {3/2}^{-} }$ are ${ \ket{ {0}^{+} } \otimes \ket{ {p}_{3/2} } }$ and 
${ \ket{ {2}^{+} } \otimes \ket{ {f}_{7/2} } }$.
The ${ {J}^{ \pi } = {9/2}^{-} }$ and ${ {11/2}^{-} }$ states are largely dominated by the ${ \ket{ {2}^{+} } \otimes \ket{ {f}_{7/2} } }$ channel (60-80\%, see Tab.~\ref{tab_defWS_channel_norms}), and are narrow resonances. These states are somehow analogous to the narrow rotational resonances predicted in ${ {}^{11}\text{Be} }$ \cite{fossez16_1335}.

As seen in Fig.~\ref{fig_2}, the g.s. assignment consistently predicted by all configuration-interaction models employed in this study is ${ {J}^{ \pi } = {7/2}^{-} }$ while the position of the ${ {J}^{ \pi } = {3/2}^{-} }$ level is strongly model-dependent.
To understand the difference between the PRM and CI predictions for this state,
we carried out three sets of GSM calculations by steadily including more continuum degrees of freedom.
	
The case indicated as GSM(a) in Fig.~\ref{fig_2} corresponds to the SM variant supplemented by the addition of the ${ {p}_{3/2} }$ continuum discretized with 15 points. The case marked GSM(b) includes 6 additional HO shells for each of the ${ {f}_{7/2} }$, ${ {p}_{1/2} }$ and ${ {f}_{5/2} }$ partial waves on top of GSM(a). Finally, GSM(c) includes the ${ 0{f}_{7/2} }$ and ${ 1{p}_{3/2} }$ neutron poles as well as their associated continua, each discretized with 15 points. It appears clearly that, for the considered model space in SM/GSM, only the ${ {f}_{7/2} }$ and ${ {p}_{3/2} }$ partial waves and their related continua are important in the description of the ${ {J}^{ \pi } = {3/2}^{-} }$ state, since the GSM(c) energy matches the RGM energy.
In fact, we checked that the RGM result for the ${ {J}^{ \pi } = {3/2}^{-} }$ state remains unchanged if only the ${ {f}_{7/2} }$ and ${ {p}_{3/2} }$ states and related continua are considered.

\begin{figure}[htb]
\includegraphics[width=0.90\linewidth]{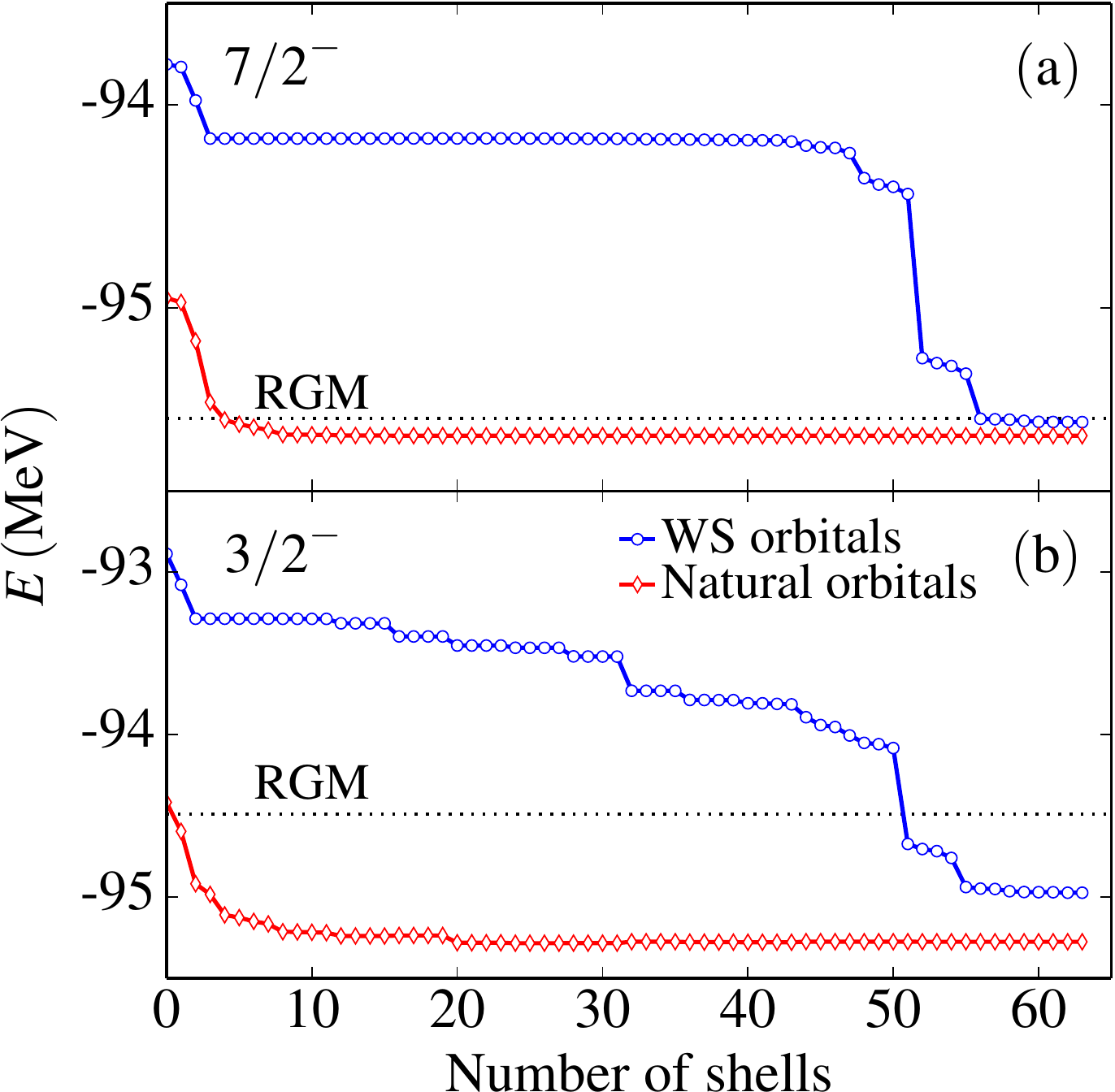}
\caption{Convergence of the DMRG energies for (a) ${ {J}^{ \pi } = {7/2}^{-} }$ and (b) ${ {3/2}^{-} }$ states in ${ {}^{39}\text{Mg} }$ with respect to the number of included shells. 
The use of natural orbitals greatly improves the convergence and provides more consistent results when compared to the traditional DMRG calculation in WS orbitals. The inclusion of additional particle-hole excitations greatly improves the description of the ${ {J}^{ \pi } = {3/2}^{-} }$ excited state, while it has little impact on the ${ {J}^{ \pi } = {7/2}^{-} }$ g.s. The DMRG convergence criterion was assumed to be ${ \varepsilon = {10}^{-8} }$. The RGM results are marked by a dotted line.
}
\label{fig_4}
\end{figure}

A significantly larger configuration space was employed in the DMRG variant.
Here, the reference space consisted of the ${ 0{f}_{7/2} }$ and ${ 1{p}_{3/2} }$ neutron shells and the ${ 1{d}_{5/2} }$ proton shell.
The complement space was made of the higher-lying neutron and proton shells: $\nu(1{p}_{1/2}, 0{f}_{5/2})$ and $\pi(1{s}_{1/2}, 0{d}_{3/2})$, respectively, as well as the corresponding continua, each discretized with 15 points.
No truncation on particle-hole excitations within the resonant space was imposed; however, only one neutron was allowed to occupy non-resonant states.
The effect of the increased model space in the DMRG is shown in Fig.~\ref{fig_4} 
for the ${ {J}^{ \pi } = {7/2}^{-} }$ and ${ {3/2}^{-} }$ states. 

It is seen that the DMRG prediction for the ${ {J}^{ \pi } = {7/2}^{-} }$ level is very close to the RGM result. This is perhaps not surprising as this state is dominated by the ${ \ket{ {0}^{+} } \otimes \ket{ {f}_{7/2} } }$ channel; hence, collective effects have little impact on it. On the other hand, a large deviation appears for the ${ {J}^{ \pi } = {3/2}^{-} }$ state, which indicates the presence of large contributions from other channels related to the quadrupole collectivity and continuum coupling. Interestingly, the DMRG energy for this state is not as low as the PRM prediction. This is presumably because the antisymmetry between the core and the valence neutron is treated approximately in the PRM as discussed below. 

The PRM prediction that the ${ {f}_{7/2} }$ and ${ {p}_{3/2} }$ partial waves are crucial for the structure of ${ {J}^{ \pi } = {3/2}^{-} }$ and ${ {7/2}^{-} }$ states is confirmed by CI models. Since in a SM picture of ${ {}^{39}\text{Mg} }$ most neutrons occupy the ${ 0{f}_{7/2} }$ shell, the sum of the weights of configurations having a fixed number of neutrons in ${ 0{f}_{7/2} }$ gives an indication on the role of s.p. states above the ${ {}^{38}\text{Mg} }$ core. The configuration weights are shown in Table~\ref{tab_config_weights} for the SM and GSM(c) calculations.
\begin{table}[htb]
	\caption{SM and GSM(c) weights of configurations (in percent) corresponding to six ${ 0{f}_{7/2} }$ neutrons and one neutron in a higher shell, 
	and also for configurations with six and seven ${ 0{f}_{7/2} }$ neutrons. The SM weights for the ${ {J}^{ \pi } = {1/2}^{-} }$ and ${ {5/2}^{-} }$ states are similar to the ${ {J}^{ \pi } = {3/2}^{-} }$ results. The ${ 0{f}_{5/2} }$ and ${ 0{p}_{1/2} }$ shells are not included in GSM(c).
		}
		\begin{ruledtabular}
			\begin{tabular}{lcccc}
			\multirow{2}{*}	{Configuration} 
			& \multicolumn{2}{c}{SM} & \multicolumn{2}{c}{GSM(c)} \\
			& ${ {3/2}^{-} }$ & ${ {7/2}^{-} }$ & ${ {3/2}^{-} }$ & ${ {7/2}^{-} }$ \\
				\hline \\[-6pt]
				${ {( 0{f}_{7/2} )}^{6} 1{p}_{3/2} }$ 	& 9 & 1 & 96 & 1 \\
				${ {( 0{f}_{7/2} )}^{6} 1{p}_{1/2} }$ 	& 0 & 0 & - & - \\
				${ {( 0{f}_{7/2} )}^{6} 0{f}_{5/2} }$ 	& 1 & 1 & - & - \\
				\\[-6pt]
				${ {( 0{f}_{7/2} )}^{6} }$		& 19 & 12 & 96 & 0 \\
				${ {( 0{f}_{7/2} )}^{7} }$		& 80 & 87 & 1 & 95 
			\end{tabular}
		\end{ruledtabular}
		\label{tab_config_weights}
	\end{table}

The dominant SM configurations are those with 7 neutrons in the ${ 0{f}_{7/2} }$ shell. The addition of the ${ {f}_{7/2} }$ and ${ {p}_{3/2} }$ continua in GSM(c) leads to a drastic change for the ${ {J}^{ \pi } = {3/2}^{-} }$ state, which becomes dominated by the ${ {( 0{f}_{7/2} )}^{6} 1{p}_{3/2} }$ configuration. While the occupation of the ${ 1{p}_{3/2} }$ shell is strongly disfavored in the SM because of its high excitation energy, the presence of the ${ {p}_{3/2} }$ continuum in GSM(c) gives rise to a structure closer to the ${ \ket{ {0}^{+} } \otimes \ket{ {p}_{3/2} } }$ channel in the PRM, which corresponds to a larger occupation of the ${ 1{p}_{3/2} }$ shell. It now becomes apparent that the high excitation of the ${ {J}^{ \pi } = {3/2}^{-} }$ SM level seen in Fig.~\ref{fig_2} is primarily due to the absence of continuum correlations. This result illustrates the failure of the traditional closed-system  SM framework when it comes to nuclear excitations above the particle threshold, involving low-$\ell$ configurations.
	
Some part of the discrepancy between the PRM and the DMRG for the ${ {J}^{ \pi } = {3/2}^{-} }$ state can be attributed to antisymmetry. 
Indeed, in CI models that include the one-neutron continuum, the ${ 0{f}_{7/2} }$ shell is largely occupied; hence, the Pauli principle reduces even further the weight of the ${ { ( 0{f}_{7/2} ) }^{7} }$ configuration with respect to the ${ { ( 0{f}_{7/2} ) }^{6} 1{p}_{3/2} }$ configuration, 
which is strongly influenced by the continuum. 
This effect is absent in the PRM picture; it should result in a larger contribution from the ${ \ket{ {0}^{+} } \otimes \ket{ {p}_{3/2} } }$ channel.

\subsection{Lifetimes}\label{sec_gs_prop}
	
Even though our CI calculations suggest a ${ {J}^{ \pi } = {7/2}^{-} }$ assignment for the g.s. of ${ {}^{39}\text{Mg} }$, 
those results are based on an interaction optimized at the SM level; hence, the addition of neutron continuum in the fit may lead to a lower ${ {3/2}^{-} }$ state.
Moreover, the targeted binding energies for ${ A > 36 }$ magnesium isotopes have been taken from systematic trends, which results in a considerable uncertainty.
Consequently, in this section we discuss the two alternative spin assignments for the g.s. of ${ {}^{39}\text{Mg} }$.

We begin with a ${ {J}^{ \pi } = {7/2}^{-} }$ g.s. scenario.
The g.s. energy of ${ {}^{39}\text{Mg} }$ (${ E \approx 130 \, \text{keV} }$) has been fitted independently in CI and the PRM, which allows us to compute the decay width and lifetime of the corresponding resonance.
The one-neutron continuum can be treated precisely within the RGM wherein
the antisymmetrization between the valence neutron (projectile) and the ${ {}^{38}\text{Mg} }$ core  (target) is properly accounted for.
The predicted width of the narrow ${ {7/2}^{-} }$ resonance in the RGM is ${ \Gamma \approx 1.2 \, \text{eV} }$, which is not too far from the PRM value of ${ \Gamma \approx 8.9 \, \text{eV} }$.

\begin{figure}[htb]
		\includegraphics[width=0.90\linewidth]{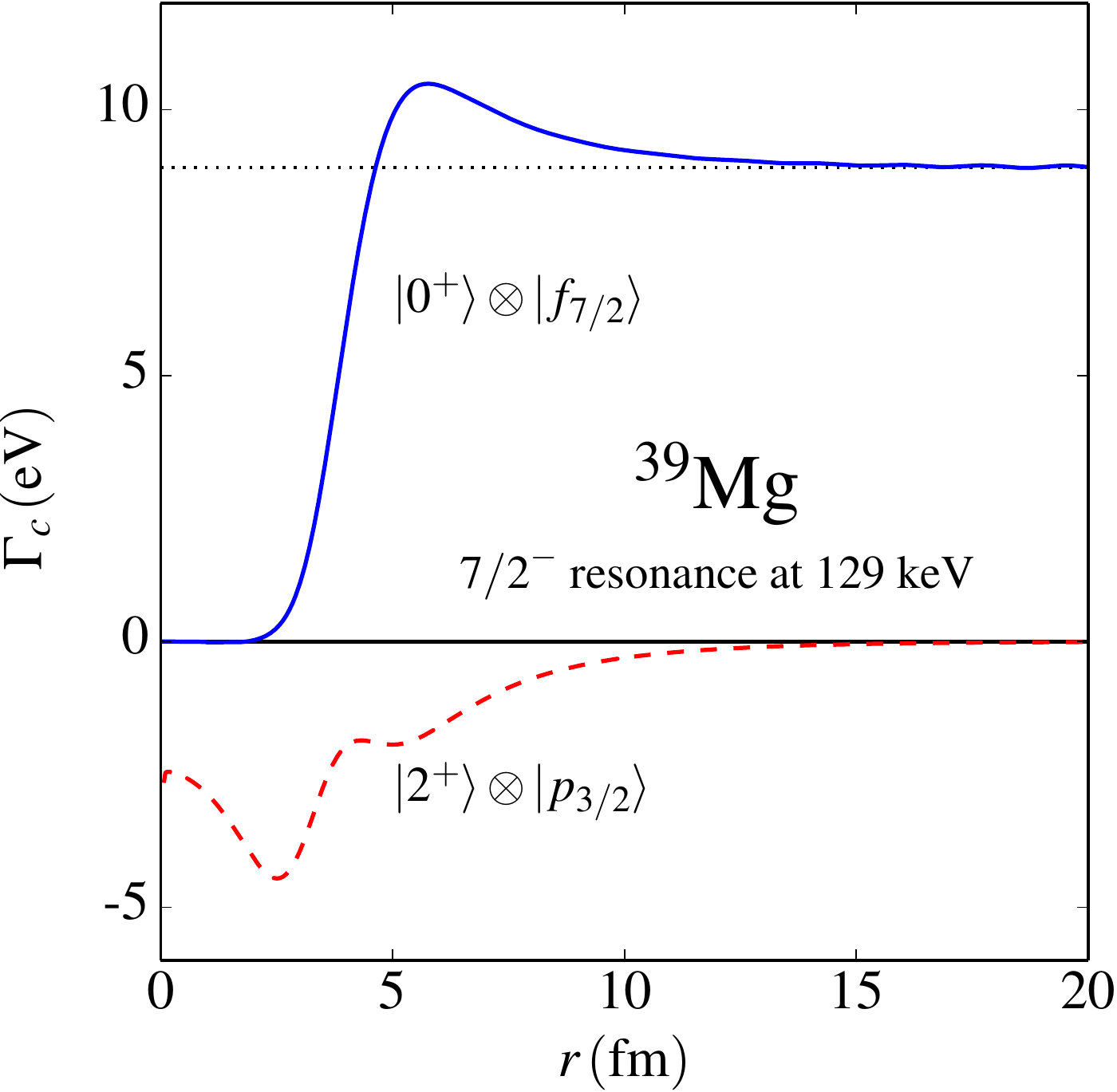}
		\caption{Partial-wave contributions of the ${ \ket{ {0}^{+} } \otimes \ket{ {f}_{7/2} } }$ and 
		${ \ket{ {2}^{+} } \otimes \ket{ {p}_{3/2} } }$ PRM channels to the decay width of the ${ {J}^{ \pi } = {7/2}^{-} }$ resonance at ${ 129 \, \text{keV} }$ in ${ {}^{39}\text{Mg} }$. The width corresponding to the imaginary part of the complex energy eigenvalue obtained by the direct diagonalization of the coupled-channel PRM equations is indicated by a dotted line.}
		\label{fig_5}
	\end{figure}

An advantage of the coupled-channel formalism is the possibility to determine the preferred decay channels of the system 
using the so-called current expression \cite{humblet61_174,barmore00_582,kruppa04_472}:
	\begin{equation}
		{ \Gamma }_{c} (r) = 
		-\frac{ { \hbar }^{2} }{ \mu } 
		\frac{ \text{Im} \left[ {u}_{c}^{ \prime } (r) {u}^{*}_{c} (r) \right] }{ \sum_{ c' } \int_{0}^{r} dr' \, { | {u}_{ c' } ( r' ) | }^{2} },
		\label{eq_channel_width_current}
	\end{equation}
	where the channel wave functions are denoted by ${ {u}_{c} (r) \equiv {u}_{c}^{ {J}^{ \pi } } (r) }$ and the total decay width is ${ \Gamma = \sum_{c} { \Gamma }_{c} (r) }$. Although values of ${ { \Gamma }_{c} (r) }$ depend on ${ r }$ in the internal region where the coupling potential terms are not negligible, the total width ${ \Gamma }$ is independent of ${ r }$, which reflects the flux conservation. 
	Figure~\ref{fig_5} shows ${ { \Gamma }_{c} (r) }$ for the ${ {J}^{ \pi } = {7/2}^{-} }$ PRM state. According to Table~\ref{tab_defWS_channel_norms}, the two dominant channels are
	${ \ket{ {0}^{+} } \otimes \ket{ {f}_{7/2} } }$ and 
	${ \ket{ {2}^{+} } \otimes \ket{ {p}_{3/2} } }$. Since the resonance's energy is less than ${ E ({2}^{+}_{1}) = 0.656 \, \text{MeV} }$ in ${ {}^{38}\text{Mg} }$, the decay through the latter channel is blocked, as evidenced by its vanishing contribution to ${ { \Gamma }_{c} (r) }$ at large distances. 
	The small width of the ${ {J}^{ \pi } = {7/2}^{-} }$ resonance is expected due its threshold character and 	
large ${ \ell = 3 }$ value.
This result is consistent with the dominance of the ${ {( 0{f}_{7/2} )}^{7} }$ configurations in GSM as shown in Table~\ref{tab_config_weights}.

\begin{figure}[htb]
		\includegraphics[width=0.90\linewidth]{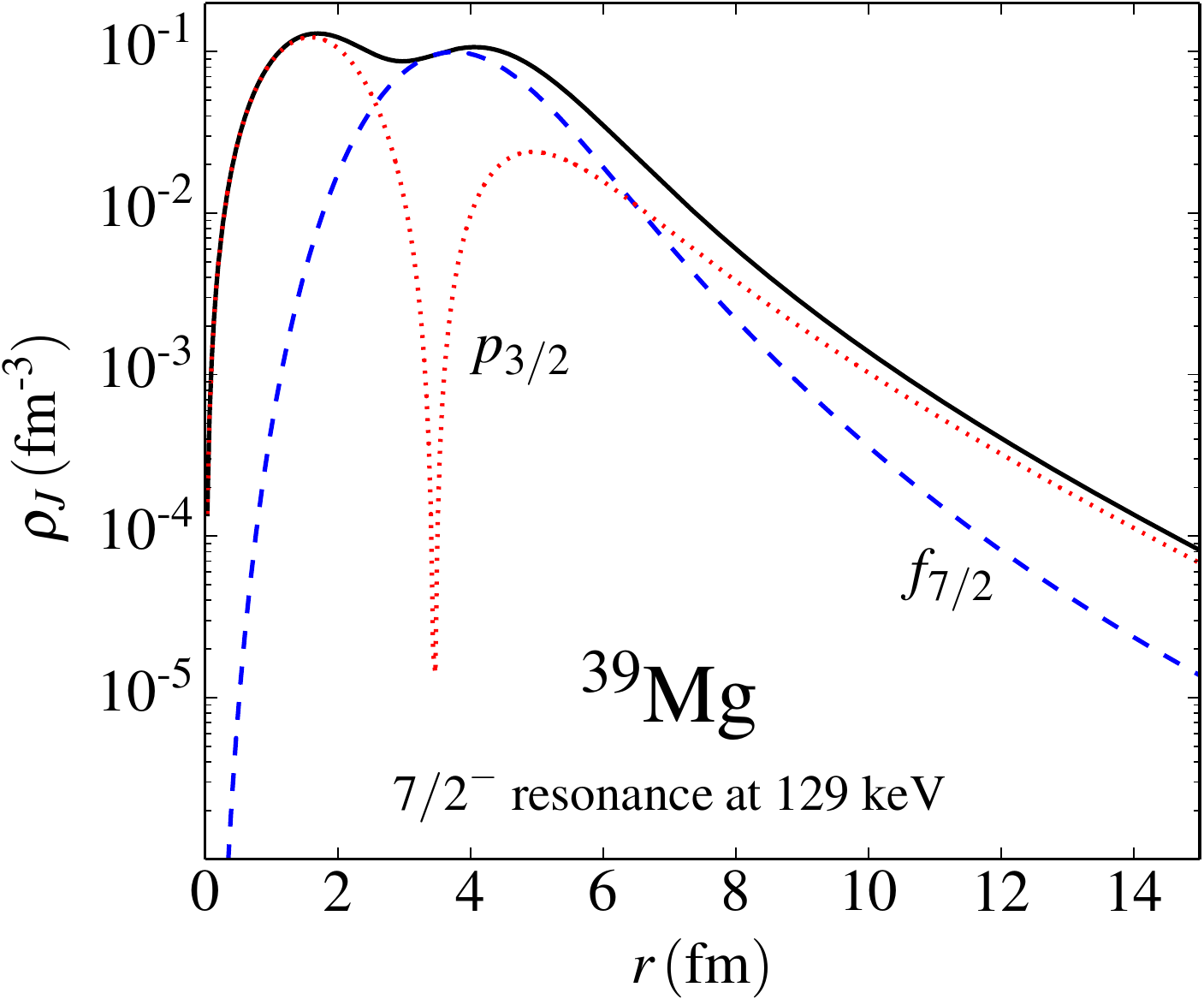}
		\caption{One-body radial density of the valence neutron of the ${ {J}^{ \pi } = {7/2}^{-} }$ resonance at ${ 129 \, \text{keV} }$ in ${ {}^{39}\text{Mg} }$ in the PRM.
		The dashed and dotted lines represent the ${ {f}_{7/2} }$ and ${ {p}_{3/2} }$ contributions, respectively.}
		\label{fig_6}
	\end{figure}

It is instructive to inspect the radial density of the valence neutron in the ${ {J}^{ \pi } = {7/2}^{-} }$ resonance. Figure~\ref{fig_6} shows the PRM result.
	It is seen that the ${ {p}_{3/2} }$ component dominates at large distances. Such a behavior is reminiscent of deformed halo systems \cite{misu97_1181} whose asymptotic behavior is governed by low-${ \ell }$ partial waves. Since the neutron-bound ${ {}^{37}\text{Mg} }$ is expected to be a one-neutron halo \cite{kobayashi14_1446,takechi14_1448} due to its
	appreciable ${ p }$-wave neutron single-particle strength, a low-lying ${ {J}^{ \pi } = {7/2}^{-} }$ resonance in 
	${ {}^{39}\text{Mg} }$ can be considered as its structural unbound analog.
	Our RGM calculations for ${ {}^{37}\text{Mg} }$ support the halo character of this nucleus,
	with the two weakly-bound ${ {J}^{ \pi } = {5/2}^{-} }$ and ${ {7/2}^{-} }$ states being the candidates for the ground state, and an excited ${ {3/2}^{-} }$ state around the threshold.

	The neutron widths of ${ J < 7/2 }$ states in ${ {}^{39}\text{Mg} }$ can be reliably predicted in the RGM. Since all those states are strongly affected by the ${ p }$-wave continuum, large widths over ${ 1 \, \text{MeV} }$ are expected, see Fig.~\,\ref{fig_2}.

We emphasize that the predicted lifetimes of the ${ {J}^{ \pi } = {3/2}^{-} }$ and ${ {7/2}^{-} }$ g.s. candidates vary by many orders of magnitude. 
The ${ \ell = 1 }$ continuum strongly impacts the structure of the ${ {3/2}^{-} }$ state and gives rise to a large width of about ${ 2 \, \text{MeV} }$.
In contrast, the ${ {7/2}^{-} }$ state is predicted to be a narrow resonance with a lifetime of about ${ {10}^{-17} \, \text{s} }$. Consequently, an experimental assessment of the width of the g.s. of ${ {}^{39}\text{Mg} }$ is critical for making a definite g.s. spin assignment.

\section{Conclusions}\label{sec_conclusion}

The low-energy structure of the unbound nucleus ${ {}^{39}\text{Mg} }$ was investigated by means of several models of different fidelity. We used the quantified interactions optimized to the binding energies of neutron-rich Mg isotopes and ${ {2}^{+} }$ excitations of ${ {}^{34,36,38}\text{Mg} }$. 
While the large quadrupole deformations ${ { \beta }_{2} = 0.3 }$ predicted by the Hartree-Fock-Bogoliubov calculations represent a challenge to some of our CI models (SM, GSM and RGM) due to configuration-space limitations,
the DMRG approach nicely connects these calculations with the PRM by significantly increasing the size of the model space. Except for the SM, all our approaches are based on the complex-energy open quantum system framework and hence can be used to describe the structure of weakly-bound or unbound states.

In spite of interaction uncertainties, largely related to the unknown masses of ${ A > 37 }$ magnesium isotopes and the  intrinsic limitations of our models (configuration-space truncations in CI approaches and a very approximate treatment of the Pauli principle in the PRM), the achieved consistency between the DMRG and PRM pictures make us believe that our conclusions about the structure of ${ {}^{39}\text{Mg} }$ are robust. As far as the g.s. structure of ${ {}^{39}\text{Mg} }$ is concerned, we predict two candidates that lie within the current theoretical uncertainty.
The ${ {J}^{ \pi } = {7/2}^{-} }$ narrow resonance width ${ \Gamma \approx 8 \, \text{eV} }$ in the PRM is expected to have a structure dominated by the 
${ \ket{ {0}^{+} } \otimes \ket{ {f}_{7/2} } }$ and 
${ \ket{ {2}^{+} } \otimes \ket{ {p}_{3/2} } }$ channels. Because the ${ \ell = 1 }$ partial wave dominates at large distances, this state can be viewed as an unbound analog of the neutron halo g.s. of ${ {}^{37}\text{Mg} }$.
The other g.s. candidate, a broad ${ {J}^{ \pi } = {3/2}^{-} }$ resonance, is expected to have structure dominated by the ${ \ket{ {0}^{+} } \otimes \ket{ {p}_{3/2} } }$ and 
${ \ket{ {2}^{+} } \otimes \ket{ {f}_{7/2} } }$ PRM channels. This state can be associated with the ${ {1/2}^{-} [321] }$ deformed Nilsson orbital.
Our results. therefore, do not support the ${ {J}^{ \pi } = {5/2}^{-} }$ g.s. assignment for ${ {}^{39}\text{Mg} }$ proposed in Refs.~\cite{hamamoto09_1452,hamamoto16_1609} but are not inconsistent with the analysis of Ref.~\cite{takechi14_1448} where a ${ {1/2}^{-} [321] }$ scenario was suggested.

The wave functions of the ${ {J}^{ \pi } = {9/2}^{-} }$ and ${ {11/2}^{-} }$ members of the g.s. rotational band are dominated by the ${ {f}_{7/2} }$ partial wave. These narrow resonances, for which the 
${ \ell = 1 }$ neutron decay is blocked, are similar to the rotational structures embedded in the continuum predicted for ${ {}^{11}\text{Be} }$ \cite{fossez16_1335}.

In conclusion, we demonstrated how the intricate interplay between single-particle motion and collective rotation, affected by the coupling to the particle continuum, 
impacts the low-energy structure of ${ {}^{39}\text{Mg} }$. Future experimental work, in particular decay-width measurements, will hopefully discriminate between the two g.s. scenarios proposed in this study.


\begin{acknowledgments}
Useful discussions with Augusto Macchiavelli and Heather Crawford are gratefully acknowledged.
We are grateful to Erik Olsen for carefully reading the manuscript.
This work was supported by the U.S.\ Department of Energy, Office of
Science, Office of Nuclear Physics under award numbers 
DE-SC0013365 (Michigan State University) and DE-SC0008511 (NUCLEI SciDAC-3 collaboration), and by the National Science Foundation under award number PHY-1403906.
An award of computer time was provided by the Institute for Cyber-Enabled Research at Michigan State University and by Chalmers Centre for Computational Science and Engineering (C3SE) through the Swedish National Infrastructure for Computing (SNIC).
\end{acknowledgments}


%

\end{document}